\documentclass[journal,twoside,web]{ieeetran}
\usepackage{cite}
\usepackage{amsmath,amssymb,amsfonts}
\usepackage{algorithmic}
\usepackage{graphicx}
\graphicspath{{../}} 
\usepackage{textcomp}
\usepackage{wrapfig}
\usepackage{xspace}
\RequirePackage[table]{xcolor}

\def\BibTeX{{\rm B\kern-.05em{\sc i\kern-.025em b}\kern-.08em
    T\kern-.1667em\lower.7ex\hbox{E}\kern-.125emX}}
\definecolor{abstractbg}{rgb}{0.89804,0.94510,0.83137}
\newcommand{\noind}[0]{\noindent}
\newcommand{\noindpar}[1]{\noind {\bf #1}}
\newcommand{\sysname}{TRAP\xspace}

\begin{document}
\title{Reliable Transiently-Powered Communication}
\author{Alessandro~Torrisi,~\IEEEmembership{Student Member,~IEEE}, 
Kas{\i}m Sinan Y{\i}ld{\i}r{\i}m,~\IEEEmembership{Member,~IEEE,} \\ and 
Davide~Brunelli,~\IEEEmembership{Senior Member,~IEEE}%   
\thanks{The authors would like to thank Dr. Przemys{\l}aw Pawe{\l}czak (Delft University of Technology) for providing with the backscatter receiver frontend boards presented in~\cite{majid2019multi}.}% 
\thanks{This work was supported by the Italian Ministry for University and Research (MUR) under the program “Dipartimenti di Eccellenza (2018-2022)”.}% 
\thanks{Alessandro~Torrisi and Davide~Brunelli, are with the Department of Industrial Engineering, University of Trento, Via Sommarive, 9, Povo, 38123 TN, Italy. e-mail: \{alessandro.torrisi, davide.brunelli\}@unitn.it}%  <-this % stops a space
\thanks{Kas{\i}m Sinan Y{\i}ld{\i}r{\i}m is with the Department of Information Engineering and Computer Science, University of Trento, Via Sommarive, 9, Povo, 38123 TN, Italy. e-mail: kasimsinan.yildirim@unitn.it}% <-this % stops a space
\thanks{Alessandro~Torrisi is also with the National Interuniversity Consortium of Materials Science and Technology INSTM, Trento research unit, Povo, 38123 TN}%  <-this % stops a space
\thanks{Corresponding author: alessandro.torrisi@unitn.it}
\thanks{This article has been accepted for publication in IEEE Sensors Journal.}
}

%\thanks{This paragraph of the first footnote will contain the date on }
% which you submitted your paper for review. It will also contain support 
% information, including sponsor and financial support acknowledgment. For 
% example, ``This work was supported in part by the U.S. Department of 
% Commerce under Grant BS123456.'' }
% \thanks{The next few paragraphs should contain 
% the authors' current affiliations, including current address and e-mail. For 
% example, F. A. Author is with the National Institute of Standards and 
% Technology, Boulder, CO 80305 USA (e-mail: author@boulder.nist.gov). }
% \thanks{S. B. Author, Jr., was with Rice University, Houston, TX 77005 USA. He is 
% now with the Department of Physics, Colorado State University, Fort Collins, 
% CO 80523 USA (e-mail: author@lamar.colostate.edu).}
% \thanks{T. C. Author is with 
% the Electrical Engineering Department, University of Colorado, Boulder, CO 
% 80309 USA, on leave from the National Research Institute for Metals, 
% Tsukuba, Japan (e-mail: author@nrim.go.jp).}

\IEEEtitleabstractindextext{%
\fcolorbox{abstractbg}{abstractbg}{%
\begin{minipage}{\textwidth}%
\begin{wrapfigure}[12]{r}{2.6in}%
\includegraphics[width=2.4in]{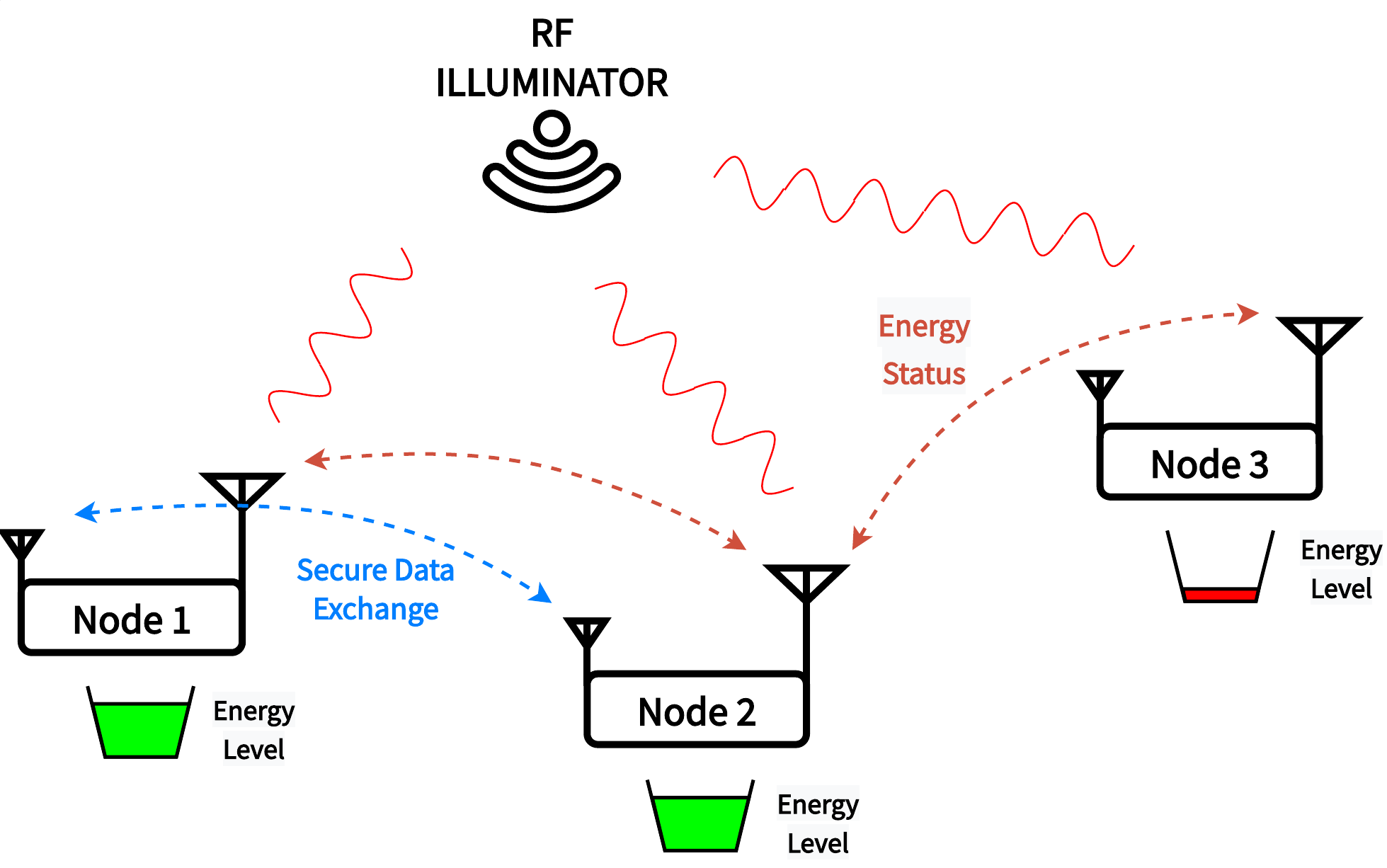}
\end{wrapfigure}
\begin{abstract} Frequent power failures can introduce significant packet losses during communication among energy harvesting batteryless wireless sensors. Nodes should be aware of the energy level of their neighbors to guarantee the success of communication and avoid wasting energy. This paper presents TRAP (TRAnsiently-powered Protocol) that allows nodes to communicate only if the energy availability on both sides of the communication channel is sufficient before packet transmission. TRAP relies on a novel modulator circuit, which operates without microcontroller intervention and transmits the energy status almost for free over the radiofrequency backscatter channel. Our experimental results showed that TRAP avoids failed transmissions introduced by the power failures and ensures reliable intermittent communication among batteryless sensors.
\end{abstract}

\end{minipage}}}

\maketitle

\section{Introduction}

\IEEEPARstart{B}{atteryless} energy harvesting devices (e.g.,~\cite{hester2017flicker,nardello2019camaroptera}) operate by relying upon several ambient sources such as solar~\cite{nardello2019camaroptera}, radiofrequency (RF)~\cite{assimonis2016harvesting, kwan2018wsn} and even bacteria species~\cite{Sartori2019}. These devices store the harvested energy in a small capacitor. They consume the stored energy conservatively to compute, sense, and communicate. When the capacitor drains out, these devices turn off due to a power failure. Therefore, the life-cycle of a batteryless device is composed of charge, sense/compute/send, and die intervals that repeat indefinitely.

Wireless communication is an indispensable requirement for batteryless sensors. Radio transmission using active radios is costly compared to the energy budget of batteryless systems~\cite{ryoo2018barnet,majid2019multi}. RF backscatter avoids the energy-hungry circuits of active radios (e.g., power-hungry mixers generating carrier waves~\cite{zhang2016enabling}), which brings almost zero-power wireless communication capabilities for batteryless nodes. In traditional RF backscatter, tags transmit by reflecting the impinging RF signals produced by a dedicated illuminator. This operation requires several orders of magnitude less energy than wireless tag transmission using active radios~\cite{Niu19overview, majid2019multi}.

\begin{figure}
    \centering
    \includegraphics[width=0.6\columnwidth]{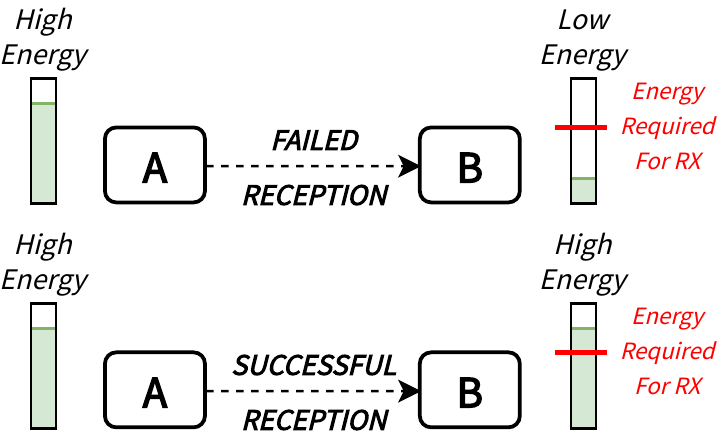}[t]
    \caption{When a high-energy transmitter device starts communication with a low-energy receiver, there might be a packet loss due to a power failure. TRAP ensures both transmitter and receiver are in a high-energy status before data transmission.}
    \label{fig:intro}
\end{figure}

The ultra-low-power wireless communication capability introduced by the RF backscatter is not sufficient to enable reliable communication among transiently-powered batteryless nodes~\cite{talla2020problems}. In particular, prior work assumed that devices are continuously powered even during zero-power communications, e.g., the RFID reader provides continuous energy.  However, batteryless nodes operate intermittently, and communication is subject to power failures. Consider the scenario between two batteryless nodes depicted in Figure~\ref{fig:intro}. Node A, which has a high energy level, wants to engage transmission with node B, which has a low energy level. In this example, the receiver node does not have enough energy to pursue the packet reception. Node B dies upon an unpredicted power failure, which leads to a packet delivery failure and wastes the precious energy spent on both sides (i.e., the transmitter and the receiver). The transiently-powered communication is successful if the energy available on both sides of the channel is sufficient to complete transmission and reception operations. %Thus, a coordination mechanism is required to guarantee reliable communication and energy efficiency.

%\noindpar{Contributions.} 
With this article, we enable reliable transiently-powered communication among batteryless devices. Our key insight is to let the transmitter device be aware of the receiver device's energy availability before data transmission. Therefore, packet transmissions happens when both nodes have sufficient energy. Briefly, we make the following contributions:

\noindpar{1-) Auto-modulator Circuit.} We designed a novel circuit to drive the backscatter circuitry in~\cite{majid2019multi}. The circuit automatically measures the energy level of the storage capacitor, encodes the measured energy level without microcontroller intervention by turning a fixed frequency, low-cost and ultra-low-power oscillator on, for a specific time. Thus, it generates a burst with different duration (i.e., number of pulses) related to the energy status.

\noindpar{2-) Transiently-Powered Communication.} We design, implement and demonstrate the first reliable communication protocol for transiently-powered devices, named as \sysname (TRAnsiently-powered Protocol). \sysname relies on the energy status information transmitted autonomously by the auto-modulator circuit over the energy status channel. % By using the energy state signal of a neighbor, a transmitter node can initiate data transmission on the data transmission channel, using a different backscatter radio dedicated to only data transmission.

A preliminary version of this work has been published in~\cite{torrisi2020zero}, which was limited to simulations. Moreover, the presented solution required microcontroller intervention for energy status transmission, which was energy-consuming for a batteryless device. We this article, we add three main extensions to the previous version. First, we design a new circuit that shares energy status information without microcontroller intervention. Second, we present a characterization of the real implementation of this circuit. Finally, we integrate our circuit to \sysname and evaluate the overall system in a real test-bed environment. Besides, we fully characterize the backscatter energy status channel to obtain a robust communication of the energy status information. 
\section{Related work}

Batteryless devices compute and communicate intermittently due to power constraints and energy discontinuity. This situation brings about several challenges. The most critical side effect of power failures is the loss of the computational state, e.g., CPU registers and memory contents clear after a power failure. This situation prevents the forward progress of computation and leads to memory inconsistencies~\cite{colin2016chain}. Therefore, existing software systems designed for continuously-powered computers fail under intermittent operation. To remedy this situation, researchers proposed checkpointing computation state of the device~\cite{balsamo2016hibernus++,kortbeek2020time}. As the computational state progresses and reaches a checkpoint, the application stores the registers and contents of the volatile memory to non-volatile memory~\cite{Balsamo2016Graceful,Arreola2015Approaches}. After a power failure, the last checkpoint acts as a backup, and the computation can progress from a consistent computational state. Another solution is to exploit task-based programming models~\cite{maeng2017alpaca,ruppel2019transactional,yildirim2018ink} offering an efficient alternative to checkpoints but requiring a non-trivial code transformation. 

Despite the progress in intermittent computing, intermittent communication has not drawn the attention of researchers yet. Current research efforts mainly targeted decreasing the energy requirements of wireless communication. Backscatter communication is an enabling technology for zero-power wireless communication within power-constrained batteryless devices. Most of the backscatter networks described in prior work including monostatic and bistatic solution~\cite{talla2017lora,hu2020ambient,alevizos2018multistatic, Jameel19healt} perform a reader-to-tag communication (i.e., a single channel exists between a dedicated master, e.g., an RFID reader, and the tag). In this approach, only the master decodes the received weak backscattered signal. Therefore, tags do not require performing complex signal processing operations, thus, allowing for simple hardware design and minimizing the energy requirements for the batteryless devices. Recent work ~\cite{liu2013ambient,ryoo2018design,ryoo2018barnet,majid2019multi} unlocked communication among batteryless devices (i.e., tag-to-tag communication). Consequently, the RF illuminator can be as simple as a single-tone carrier generator. In this case, each node becomes a transceiver including both the encoding and decoding processes. Especially, decoding has been tailored for low power applications using only low-power analog operations (e.g., a diode-capacitor envelope detector, an operational amplifier, and a comparator). Moreover, hybrid systems proposed in ~\cite{in2016blisp,rostami2018polymorphic} combine an active and a backscatter radio having both benefits of high throughput and long-distance. Especially, long-range backscatter communication has been tackled in ~\cite{varshney2017lorea} exploiting LoRa. Furthermore, researchers unlocked easy deployment by exploiting already existing ambient RF signals such as TV and Wi-Fi ~\cite{ji2019ambientback, bharadia2015backfi, liu2013ambient, yang2018ambient}. Thus, reducing complexity as the illuminator can be already present both in indoor and outdoor scenarios. Finally, backscatter communication can be combined with RF energy harvesting systems ~\cite{assimonis2016harvesting, ryoo2018barnet, Jameel19harvest, Li2020harvest, kwan2018wsn, ryoo2018design}. The RF carrier is used not only for backscattering the information but also for collecting a small quantity of energy stored in the device and used to complete small tasks. Kwan et al.~\cite{kwan2018wsn} explored the techniques to allow for transmission and harvesting optimization. As the ambient RF energy has pretty weak spectral power, the RF harvested energy is quite small compared to other sources~\cite{assimonis2016harvesting, Roy2021}. Despite all these advancements, all these prior works require the devices to operate continuously during data packet communication. However, batteryless devices operate intermittently, which will lead to a significant amount of failed data transmissions.

Batteryless devices lose the notion of time upon power failures. The time registers lose their contents after each power failure, and the batteryless device cannot measure the time elapsed during the charging period. Maintaining a continuous notion of time is a crucial requirement to implement networking protocols, e.g., to generate periodic events for data transmissions. Recent works proposed several ultra-low-power persistent timekeepers based on RC networks~\cite{hester2016persistent, yildiz2021persi, hester2016timekeeper}. These timekeepers measure time by considering the discharging characteristics of dedicated capacitors. Briefly, the dedicated capacitor charges to a specific voltage when the batteryless device is on. The capacitor slowly discharges when the device turns off. The off-time is estimated by measuring the voltage decay across the capacitor upon reboot. There are also zero-power timekeepers, as presented in~\cite{zhou2017zptimekeeper, zhou2013zpthermal}, being fabricated in standard CMOS technology and exploiting thermal-noise energy for a self-powered clock system. Our solution presented in this article requires a continuous notion of time for periodic energy status updates and synchronization. For this purpose, we used a generic off-the-shelf ultra-low-power real-time clock for its easy accessibility and integration, rather than using the mentioned custom persistent timekeeper solutions. There are also other solutions in the literature, such as relying on external energy sources (e.g., indoor light flicker~\cite{geissdoerfer2021bootstrapping}), to synchronize the batteryless nodes.
\begin{figure*}[htbp]
    \centering
    \includegraphics[width=1.9\columnwidth]{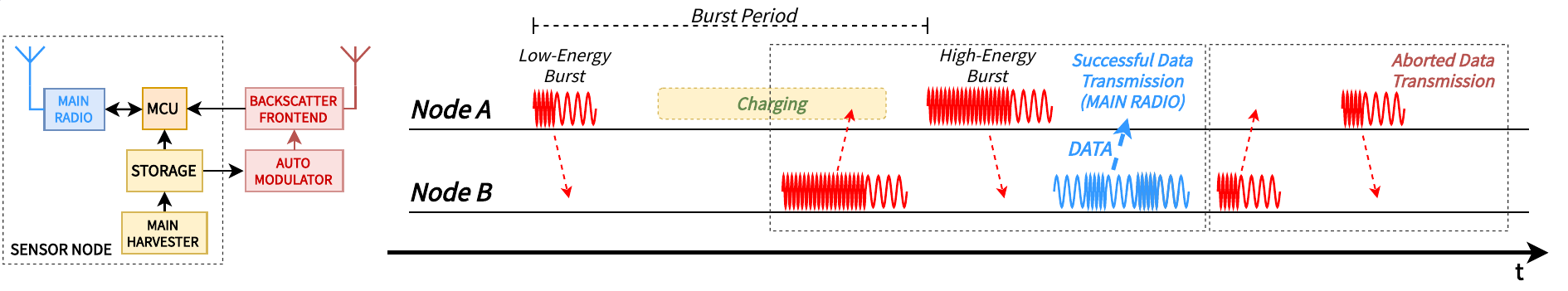}
    \caption{Example scenario with two batteryless nodes that share the energy status information over the backscatter channel.}
    \label{fig:scenario}
\end{figure*}

\section{Auto-Modulating Energy Status Information}
\label{sec:energy-status}

To perform the packet transmission and reception operations successfully, we must ensure a sufficient energy level on both sides of the communication channel. Otherwise, either the transmitter or the receiver (or both) fails, interrupting the communication. This situation will lead to significant packet losses and waste of precious harvested energy at the transmitter and receiver sides. We present a solution based on sharing \emph{energy state information} among the transiently-powered devices, which lets their neighbors decide to engage or postpone the transmission. An example scenario with two nodes is presented in Figure~\ref{fig:scenario}. Here, the energy status is shared between nodes, allowing them to either engage or abort data transmission and ensuring both sides of the communication channel have sufficient energy to conclude the operation properly.

Due to its ultra-low-power characteristics, sharing energy status information via RF backscatter is a promising approach. Exploiting an already deployed RF power source, the backscatter communication modulates the reflection of the impinging RF signal allowing for zero-power communication for the end nodes.  The main challenge we tackle is how to modulate the RF reflection to encode the energy state information avoiding power-hungry components and circuits. To this end, we present a novel circuit, named auto-modulator, that shares the energy status via modulating a burst signal based on an ultra-low-power and low-frequency oscillator, without microcontroller intervention (as opposed to our previous circuit presented in~\cite{torrisi2020zero}). 

The core of the circuit we developed, as presented in Figure~\ref{fig:block-diagram}, can be divided into two main blocks: the backscatter frontend and the low power auto-modulator. Furthermore, the microcontroller decodes the received bursts and decides to engage a reliable communication or postpone it.

\subsection{Backscatter Frontend}
\begin{figure}[b]
    \centering
    \includegraphics[width=\columnwidth]{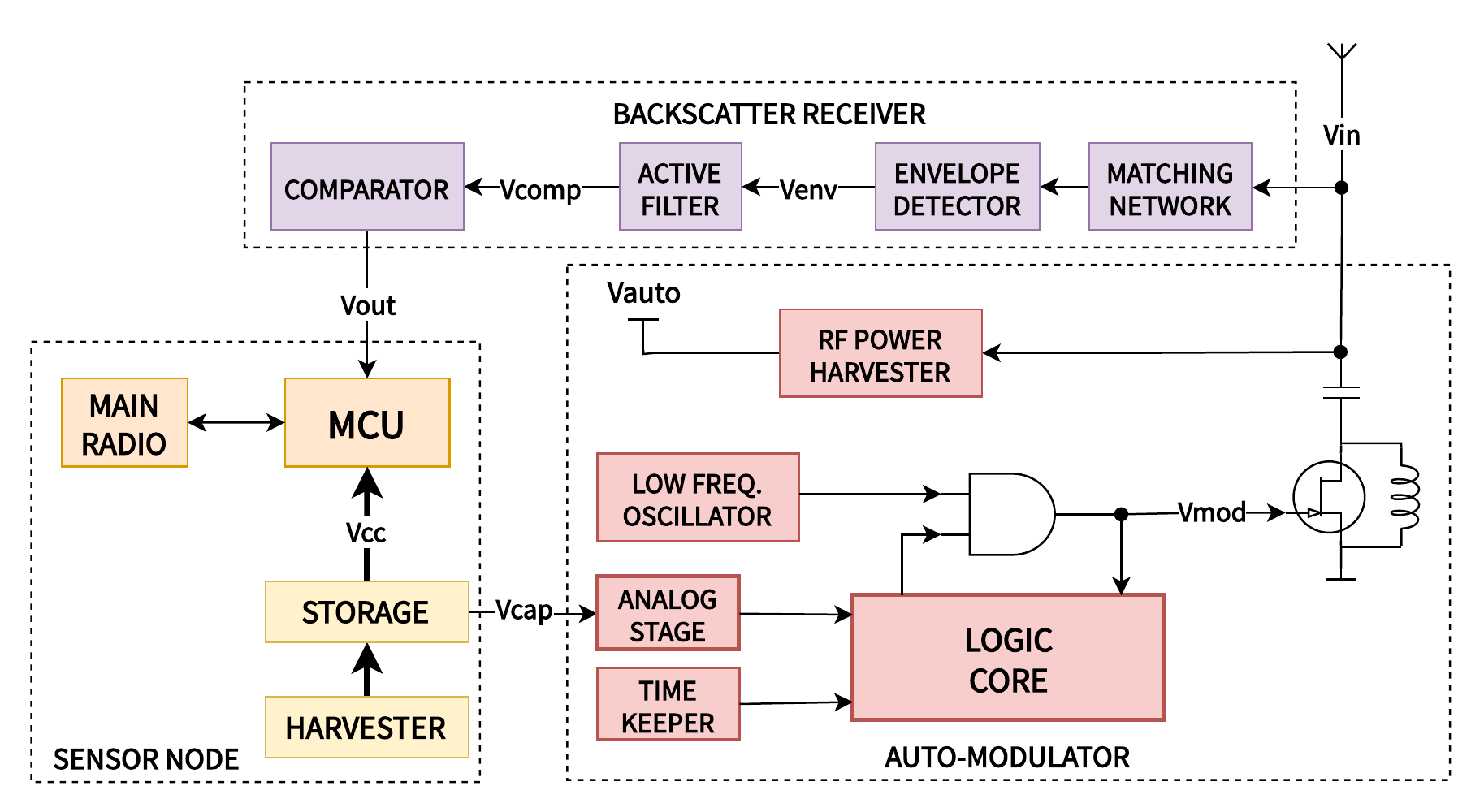}
    \caption{The auto-modulator is our novelty, providing an autonomous mechanism to share the energy status information. The front-end backscatter circuit is used as a transceiver. The MCU, when enough energy is available, decodes the information from the backscatter channel and decides either to engage or post-pone a data transmission. Harvester and energy storage are strictly dependent on the application.
    }
    \label{fig:block-diagram}
\end{figure}

%\noindpar{Backscatter frontend.} 
In backscatter communication, the transmitter device exploits different match impedances connected to RF switches to modulate the signal reflection. We exploit a single switch based on an RF MOSFET to support the ON-OFF keying (OOK) modulation. When the MOSFET is in the off state, the antenna is matched with the matched impedance presented by the receiver circuit, which absorbs the input RF signal with almost zero reflection. When the MOSFET is in the on state, the antenna is shorted out and mismatched reflecting the input RF signal. The auto-modulator continuously governs the (\texttt{Vmod} signal in Figure~\ref{fig:block-diagram}) providing the modulation. The modulation process proceeds even in the case of the node power failure as the auto-modulator relies on the RF harvested energy.

The receiver decodes the backscattered energy status information when the end node has enough energy to perform the necessary computation. We exploit the RX block presented in~\cite{majid2019multi} without any modifications. 
It includes a demodulator for simpler modulations such as OOK with lower data rates. The core is a biased Schottky diode envelope detector that performs the frequency shift in the baseband by implementing a low-power and cheap circuit. The circuit is finely matched with the RF input (\texttt{Vin}) and the antenna (see Figure~\ref{fig:block-diagram}) at 868MHz frequency. As the envelope detector output voltage swing is quite low, the following circuit is an active high pass filter stage (based on the MAX9914 IC) feeding the comparator \texttt{Vcomp}. The comparator stage (based on the LM7215 IC) performs an averaged thresholding and provides the final digital output \texttt{Vout}. The receiver node performs the energy status decoding with the MCU intervention. The MCU is interrupted by the digital output signal and it measures the burst duration and frequency, identifying the energy status of the transmitter neighboring node.

Finally, as demonstrated in~\cite{assimonis2016harvesting}, a tiny amount of energy can be harvested using the backscatter channel from the incident RF carrier or RF surrounding ambient. As described in section~\ref{sec:evaluation}, our circuit draws about 10$\mu$W. Therefore, the impinging RF power must be above -13dBm to operate our auto-modulator circuit properly, considering a cautionary RF-to-DC conversion efficiency of about 20\%. 

\subsection{Energy Status Encoding}
\label{sec:energy-status}
The energy status information is modulated in the form of a burst, i.e., a sequence of OOK modulated pulses. With this scheme, we have some knobs or encoding the energy information. Using different burst duration and the OOK modulation frequency, we can let nodes identify each other and their energy levels.
We propose four different burst durations to distinguish between four different energy levels. The lowest level (a burst of 32 pulses) indicates the node is in a charging transient, and its energy is too low to perform any tasks. The middle level (a burst of 64 pulses) indicates the node can accomplish some small tasks but its stored energy is not sufficient for reliable data transmission. The high level (a burst of 128 pulses) indicates the node is active and able to perform a reliable data reception. Finally, the highest level (a burst of 256 pulses) indicates that the energy storage is full, and the node can perform complex tasks such as computing and transmitting data.

Moreover, the burst is characterized by an OOK modulation frequency. Slightly different frequencies, in a range of about 30kHz, are used to distinguish nodes from each other. Finally, the repetition period of the burst determines the update rate of the energy status in the network. In this article, we propose a repetition period of 100ms.

\subsection{Auto-modulator.} 
The auto-modulator automatically provides the signals at the backscatter front-end encoding the energy status information discussed above (\texttt{Vmod} in Figure~\ref{fig:block-diagram}). The challenge is to run this circuit without MCU intervention, even in the case of a power failure or low energy availability on the node. Hence, it must have an ultra-low-power requirement. The following is a description of the circuit of the auto-modulator block schematized in Figure~\ref{fig:block-diagram}.
We propose to drive the modulation by a low-frequency low-power oscillator. A commercial ultra low power SiT1533AI-H4-DCC-32.768E oscillator is used, providing a frequency of 32.768kHz for the only modulation. A second ultra-low-power timer, the TPL5111, is used as a timekeeper to fire a burst transmission every 100ms. Further improvements can be done by using a zero-power persistent timekeeper~\cite{de2020reliable}.
The circuit includes an analog stage to provide the readings of the energy level coming from the storage (\texttt{Vcap}). It provides a reference voltage and three voltage thresholds, which can be set by a high-value resistors' network. We used three TLV841 ICs to obtain three different energy level thresholds. Thus, providing the four different energy levels. Finally, a logic combiner is used to produce the desired burst at the \texttt{Vmod} output, in particular, using standard CMOS technology with the CD4040BM96 binary counter main IC. Improvements can be done by using an LVCMOS standard or a VLSI implementation.

\subsection{Digital Decoding.}
\label{sec:decoding}
Thanks to the low power requirement of the auto-modulator, the backscatter channel is always populated by energy level bursts. The role of the MCU is to decode the transmitted energy levels and, in the case of energy availability, perform a reliable data transmission or reception. Specifically, the MCU takes the following steps during energy status reception:
\begin{enumerate}
    \item The backscatter front-end digital output \texttt{Vout} interrupts the microcontroller, which represents high-to-low and low-to-high bit transitions;
    \item The microcontroller accumulates the bits and forms the transmitted burst;
    \item Considering the duration of the burst, the microcontroller decodes the energy status and decides to transmit data, avoiding packet losses due to power failures.
    \item The MCU identifies the transmitter node, considering the frequency of the received and decoded burst. 
\end{enumerate}

As a final remark, the energy status reception mechanism is active only in the case of a high energy level and if the application needs to transmit sensible data. Only in this case, the MCU runs the TRAP protocol. Otherwise, all chain, including the backscatter receiver, is disabled and no interrupts are collected saving further energy.
\section{Reliable Intermittent Communication}
\label{sec:protocol}

We present \sysname (TRAnsiently-powered Protocol)~\cite{torrisi2020zero} and its fundamental building blocks in this section. \sysname ensures reliable intermittent communication by sharing the energy status information of the nodes. \sysname aims to ensure that transmitter and receiver devices have sufficient energy before starting a data packet communication. The protocol exploits the auto-modulator circuit (as presented in Section~\ref{sec:energy-status}) to extract the energy status information of the neighboring nodes. Briefly, if a node with a high energy level is detected, the data communication process can start. Otherwise, communication is postponed.

\sysname requires an illuminator that generates the necessary carrier waves for the RF backscatter communication. It is worth mentioning that the need for the dedicated illuminator can be eliminated by exploiting ambient RF signals (e.g., TV signals or Wi-Fi signals) that already exist in most indoor and outdoor  environments~\cite{liu2013ambient, bharadia2015backfi}. Also, note that our auto-modulator circuit includes a harvester to benefit from the available environmental RF energy~\cite{ryoo2018design,  ji2019ambientback, Jameel19harvest}. In this sense, the auto-modulator circuit operates {\bf without any impact} on the node's energy consumption since it operates in a self-sustainable manner by harvesting energy through its energy status backscatter channel (as depicted in Figure~\ref{fig:block-diagram}). Nonetheless, we separate the energy status channel from the data channel to ensure simultaneous data and energy status transmission and prevent collisions among data packets and energy status updates. Therefore, we can exploit a second backscatter radio for data transmission, which operates at a different frequency than the backscatter radio used for energy status transmission, or we can use another type of radio, such as an active one achieving long-distance communication.

In \sysname, each node transmits its energy availability via the auto-modulator at a \emph{fixed period}. The period is the same for all the nodes, and the auto-modulator runs indefinitely. The low power clock in the auto-modulator might introduce a possible drift, which might even be useful to prevent energy status transmission overlapping of the neighboring nodes. A node having sufficient energy to transmit data engages data packet transmission by considering the energy status information received from its neighboring node. The following steps are taken to perform reliable data transmission:

\begin{enumerate}
    \item Transmitter node checks if there is sufficient energy to perform computation and transmission. If the energy is sufficient, it starts listening to the energy status channel.
    \item During energy status reception, it decodes the energy status packet as described in~\ref{sec:decoding}. It also determines the identifier of the  neighboring node as described in~\ref{sec:energy-status}.
    \item It initiates the backscatter communication over the main radio if the neighboring node has enough energy to receive the packet.
\end{enumerate}

\noindpar{Power Failures.} The auto-modulator circuit preserves the periodicity of the energy status transmission, since its is active during the period across power failures. After a power failure, the node will harvest sufficient energy, and its energy status changes to a high level. Then, it starts transmitting a high-energy burst. Any node that detects the high-energy burst can immediately reply by sending data via the main communication channel. This communication scheme guarantees the energy availability of both sides, and packet losses due to power failures are eliminated. If any node receives a low-energy burst, it aborts the transmission of data to save energy. In summary, \sysname provides a notion of coordination among the nodes to prevent transmission failures, and in turn, energy waste.

\noindpar{Carrier Sense.} In \sysname, two nodes can simultaneously transmit a packet upon receiving a high-energy burst from a common neighbour, indicating the availability to receive data packets. This critical situation may lead to a data packet collision on the data communication channel. \sysname can further be extended to exploit CSMA (carrier-sense multiple access), a common solution to prevent packet collisions. In the case of energy availability, before engaging data packet transmission, nodes might generate a random back-off and check for any bit reception over the data communication channel (i.e., a carrier sensing approach). The node can securely start the data communication at the end of the back-off period if no bit is received.
\begin{figure}
    \centering
    \includegraphics[width=0.7\columnwidth]{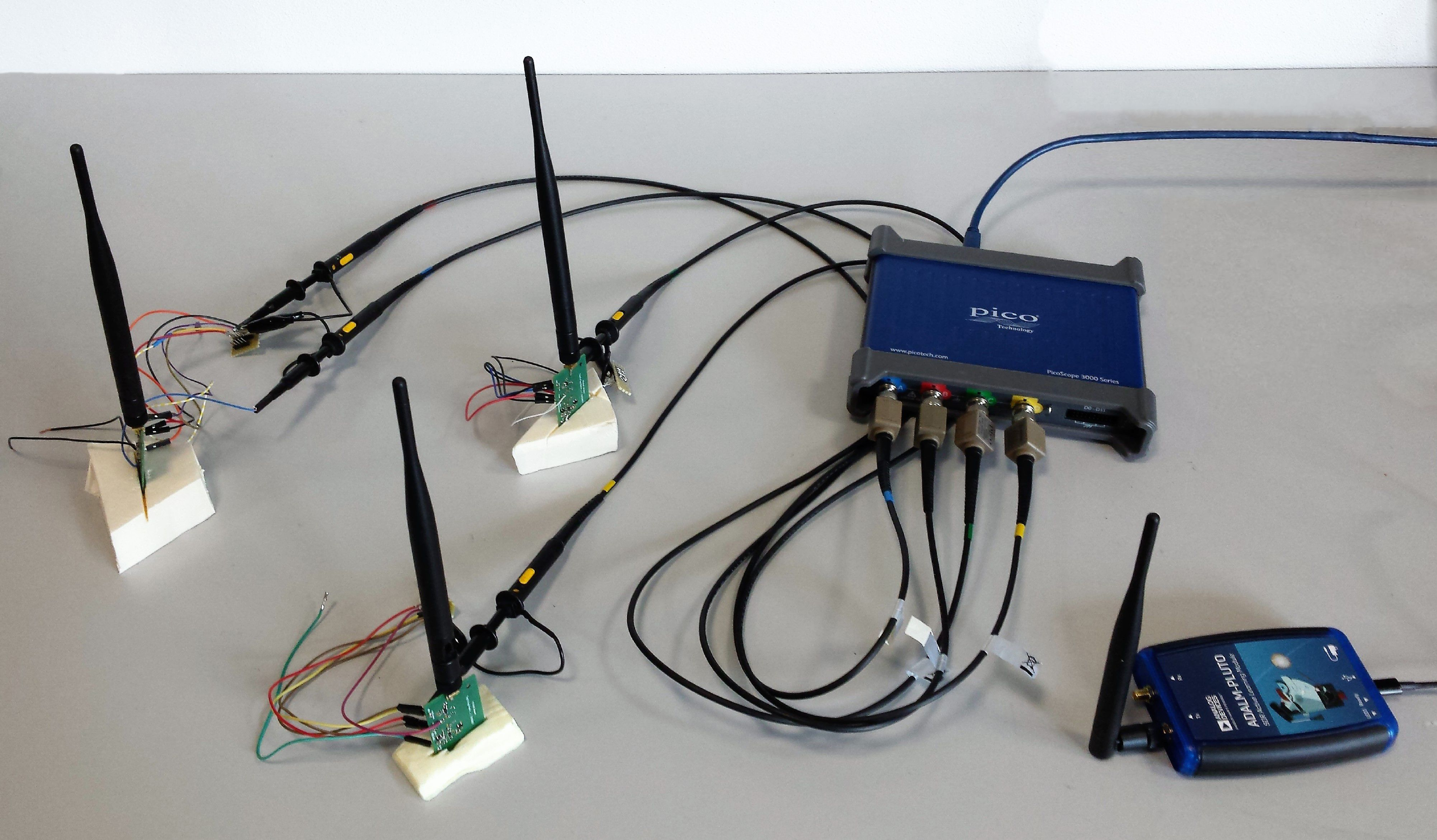}
    \caption{Our testbed setup with the three nodes, a 4 channel PicoScope 3000 and the illuminator AdalmPluto SDR}
    \label{fig:photo}
\end{figure}

\section{Evaluation}

\label{sec:evaluation}

In this section, we present our experiments and present evaluation results. We performed several measurements, including power consumption, to fully characterize the proposed energy status channel and auto-modulator circuit. Moreover, we implemented \sysname and evaluated it on a real testbed setup presented in Figure~\ref{fig:photo}. In our testbed, we used the RF backscatter frontend circuit described in~\cite{majid2019multi} for data and energy status transmission. We implemented the auto-modulator in hardware to encode the energy status. We used the Nucleo STM32L476RG board as the MCU that runs the TRAP. For measurements, we used the PicoScope 3000 as a 4 channels oscilloscope to capture the analog and digital signals, and the Adalm Pluto SDR as an illuminator that provides the RF carrier at -20dBm at 868MHz. Finally, a TextronixRSA306B spectrum analyzer is used to analyze the RF backscatter channel.
\subsection{Characterization of Auto-modulator and Backscatter}
We report an overview of the characteristics of the auto-modulator and backscatter circuits concerning the energy consumption and the main electric signals in the system. 

\noindpar{Energy Budget.} The most challenging aspect we target is to keep the auto-modulator circuit alive even in the case of low energy availability and power failure. Thus, energy consumption becomes a crucial aspect. While active for reception, the backscatter transceiver consumes up to 36$\mu$A at 2.2V roughly 80$\mu$W. Our objective is to reduce extra power consumption below this threshold. Our measurements showed the power consumption of the auto-modulator is about 10$\mu$W, well below the fixed target of 80$\mu$W. Table~\ref{tab:energy-budget} reports an overview of the power distribution in the auto-modulator.

\begin{table}
\centering
\caption{The energy budget for the auto-modulator is divided in the main sources. The supply voltage is fixed at 3.3V.}
\label{tab:energy-budget}
 \begin{tabular}{ m{3.9cm} m{1.9cm} m{1.7cm}} 
 \hline
  & {\bf Current} [$\mu$A] & {\bf Power} [$\mu$W]\\
 \hline
 {\em Logic core + Vmod Driving} & 0.39 & 1.29 \\
 \rowcolor{black!3}
 {\em Time keeper} & 0.04 & 0.13 \\
 {\em Low frequncy oscillator} & 2.06 & 6.80 \\
 \rowcolor{black!3}
 {\em Analog stage} & 0.48 & 1.58 \\
 {\em Overall power} & 2.97 & 9.80 \\
 \hline
\end{tabular}
\end{table}

In the auto-modulator circuitry, the main power-hungry components are the low-frequency ultra-low-power oscillator, drawing roughly 6.80$\mu$W, and the analog stage, including a reference voltage to establish fixed thresholds and drawing about 1.58$\mu$W. The timekeeper must always be alive; thus, the RF harvester must provide sufficient energy for its operation. In our case study, we exploited as COTS component the TPL5111 drawing up to 0.13$\mu$W. Moreover, a gating mechanism can further reduce the system power requirement (i.e., all the elements except the timekeeper) if the burst repetition period is longer than 600ms. Indeed, the settling time for the SiT oscillator is 300ms (as reported in the datasheet~\cite{sit1533}). Moreover, the same technique is used to lower the backscatter front-end average power consumption, which is on only in the case of need. Finally, we can further reduce the power consumption of the logic core by replacing the standard CMOS technology and COTS components with LVCMOS or even a VLSI implementation.

\begin{figure}
    \centering
    \includegraphics[width=0.75\columnwidth,trim={3cm 10cm 2.5cm 10.4cm},clip]{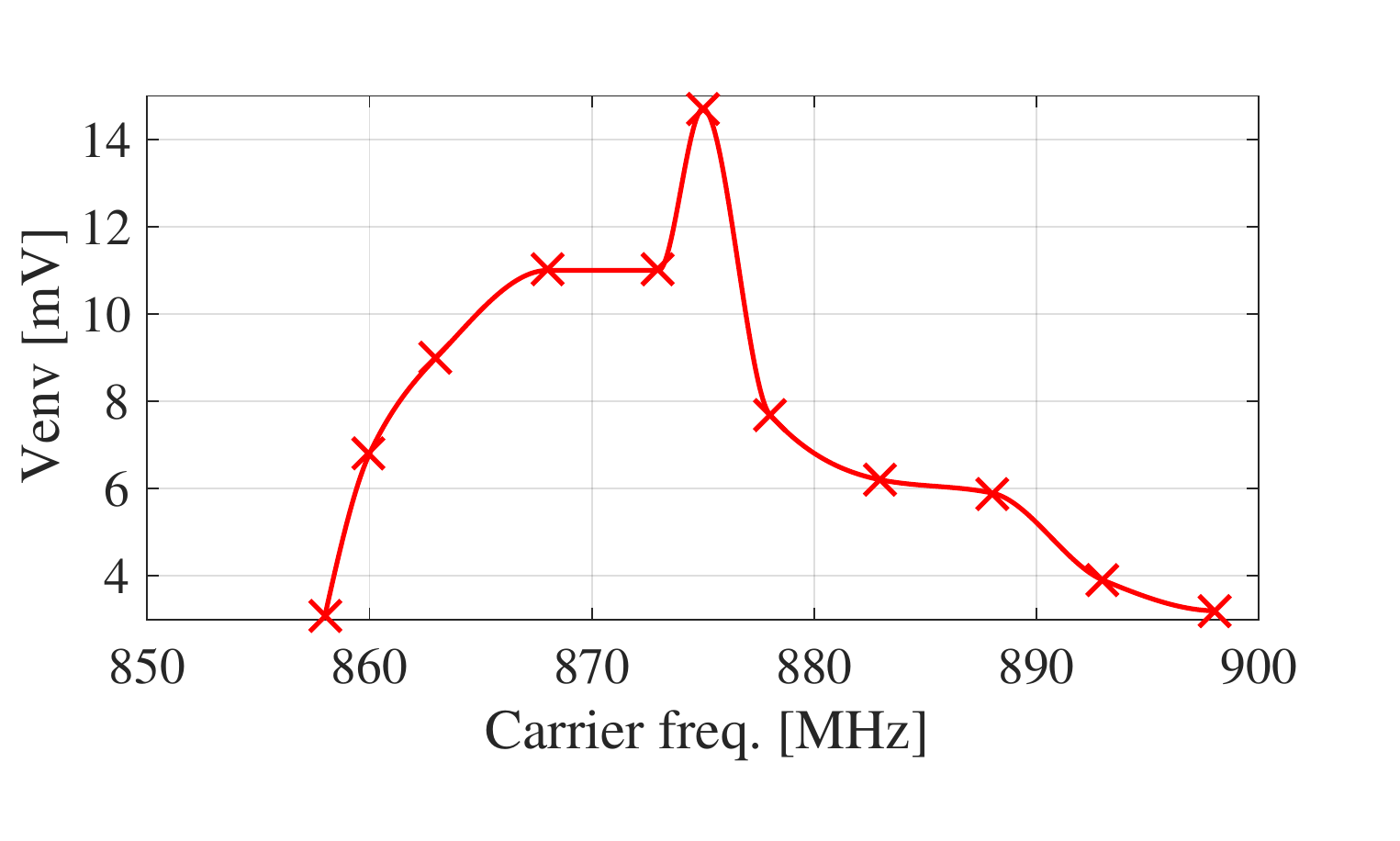}
    \caption{The peak-to-peak envelope detector voltage (\texttt{Venv}) as a function of the carrier frequency (at 868MHz center freq.}
    \label{fig:fbw}
\end{figure}

\noindpar{Burst Reception.} To verify the quality of the received signals and bursts, we built a test setup with two nodes: one is the backscatter transmitter, one is the backscatter receiver. The illuminator providing the RF carrier is an Adalm Pluto SDR set to the maximum output power. First, we measure the RF bandwidth of the transceiver circuit in the form of fractional bandwidth (FBW). It is important to understand the sensitivity of the receiver concerning the carrier frequency, as different carriers can be used to transmit both the information on the energy status and the actual data ~\cite{torrisi2020zero}. We set the illuminator at different frequencies in a range between 858MHz and 898MHz as we expect a center frequency of 868MHz. The result is shown in Fig.~\ref{fig:fbw} reporting the receiver envelope detector peak-to-peak voltage (a valid image of the received signal strength) as a function of the carrier frequency. Even if the maximum peak-to-peak appears at a frequency of 875MHz, the related envelope detector signal appears distorted, thus the choice to keep the carrier frequency at 868MHz for reference of the following measurements. Finally, the FBW is evaluated as $(f_{2} - f_{1})/f_{c} =(889-859)/868=3.5\%$ assuming the two corner frequencies at -3dB. 

\begin{figure}
    \centering
    \includegraphics[width=0.8\columnwidth, trim={3cm 8.4cm 2.5cm 8.2cm},clip]{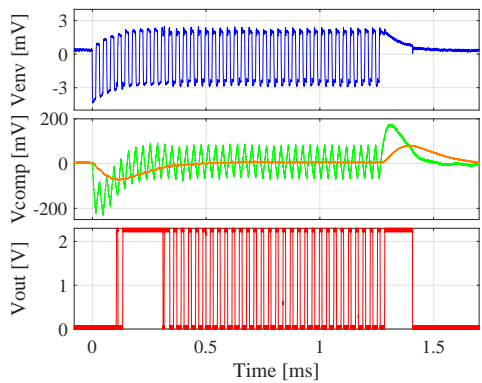}
    \caption{Representation of a 40 pulses @31kHz received burst (i.e. a burst duration of 1.3ms). The waveforms refer to the receiver side and in particular: blue trace is the envelope detector \texttt{Venv} (roughly 5mVpp); orange and green traces are the comparator inverting and non-inverting inputs \texttt{Vcomp} (AC coupled at 1.1V DC); red trace is the digital output \texttt{Vout}.}
    \label{fig:receiveddelay}
\end{figure}

Fig.~\ref{fig:receiveddelay} reports a received burst of 40 pulses at a OOK modulation frequency of 31kHz and a burst duration of 1.3ms. The four waveforms in the figure refer to:

\begin{enumerate}
    \item The envelope detector voltage at the receiver (\texttt{Venv} blue trace), with the received and demodulated burst having a very small voltage swing above 5mV peak-to-peak. It is an evidence that this very small signal needs to be conditioned before reading it with the MCU.
    \item The comparator inputs (\texttt{Vcomp} orange and green traces) are respectively the inverting  \texttt{Vcomp-} and non-inverting \texttt{Vcomp+} inputs. Thus, the inverting input is produce by low pass filtering the non-inverting one.
    \item The digital comparator output signal (\texttt{Vout} red trace). This signal is generated by the backscatter circuit on the receiver side and read by the MCU.
\end{enumerate}
 
As expected due to the signal conditioning network, a slight delay appears between the start point of the burst reception (\texttt{Venv}) and the effective digital output commutation (\texttt{Vout}) with a consequent loss of the first few pulses. Even if that can be easily managed by the MCU, considering a possible slight variation in the number of received pulses, we dig deep into this phenomenon to better understand the limit of the circuit and the possibility to differentiate even more energy levels.

As it can be seen in Fig.~\ref{fig:receiveddelay} (OOK modulation frequency of 31kHz and transmitted 40 pulses), the settling time from the beginning of the burst and the stable activation of the digital output takes 9 pulses (i.e., 0.29ms). We consider this as a hardware limitation mainly given by the receiver analog section and, in particular, the amplifier and the comparator stage. First, the envelop detector voltage (\texttt{Venv} blue trace) is amplified giving the non-inverting comparator voltage (\texttt{Vcomp+} green trace). The amplifier itself modifies the signal not only in the amplitude (roughly a gain of 30) but also in the shape due to the limited bandwidth. Secondly, a low pass filter is used to obtain, from the \texttt{Vcomp+}, the average threshold applied to the comparator inverting input (\texttt{Vcomp-} orange trace). In the first time interval after the burst beginning, \texttt{Vcomp-} does not react as fast as the \texttt{Vcomp+}. Consequently, the comparator overdrive is negative and the output remains low. When the \texttt{Vcomp-} reaches and passes the  \texttt{Vcomp+}, some pulses are properly generated, but the averaging filter is still away from the steady-state. In the following time interval and until the \texttt{Vcomp-} signal stabilizes at a middle value, the comparator is not able to distinguish the pulses due to the low overdrive. In turn, the comparator LMC7215 has a relatively large propagation delay at low input overdrive voltage.

\begin{figure}
    \centering
    \includegraphics[width=0.75\columnwidth]{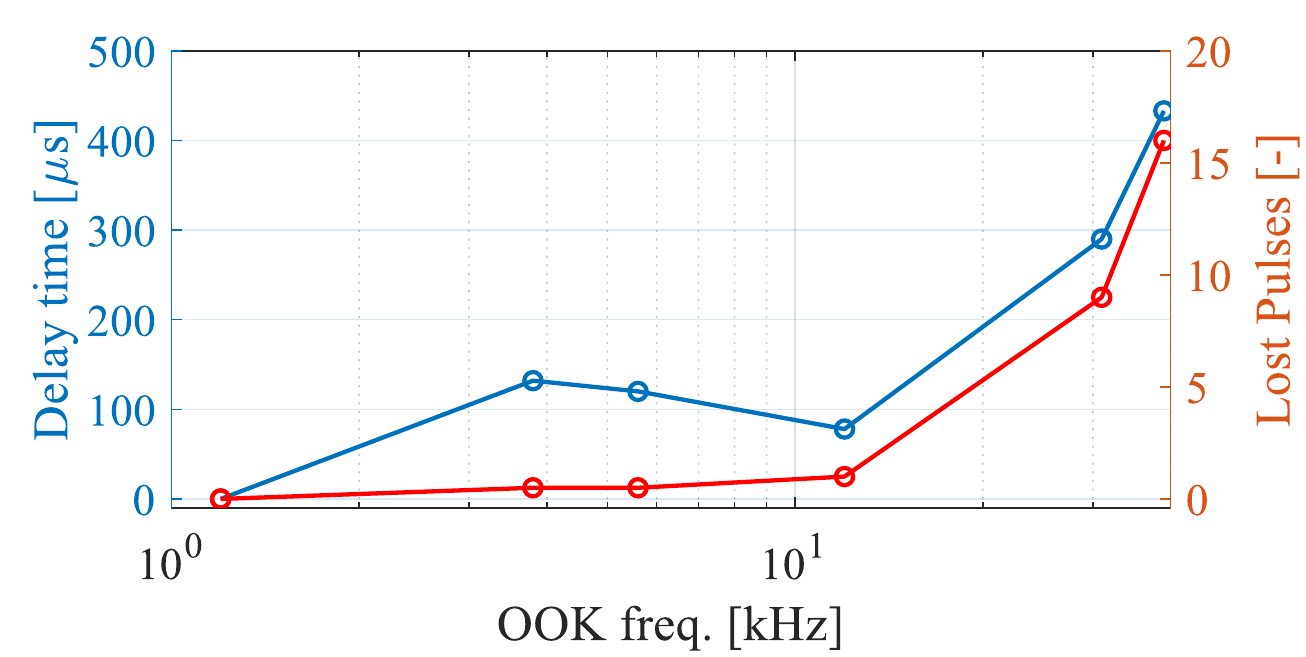}
    \caption{Representation of the settle time and number of lost pulses at the beginning of the burst reception. %(refered to Fig.~\ref{fig:receiveddelay}.
    %The delay time and the number of lost pulses are reported as a function of the OOK modulation frequency. The envelope detector peak to peak remains almost constant over the depicted range.  With increasing frequency the settle time is longer and a more pulses are lost.
    }
    \label{fig:resdelay}
\end{figure}

\begin{figure}
    \centering
    \includegraphics[width=0.75\columnwidth,trim={3cm 10cm 2.5cm 10.3cm},clip]{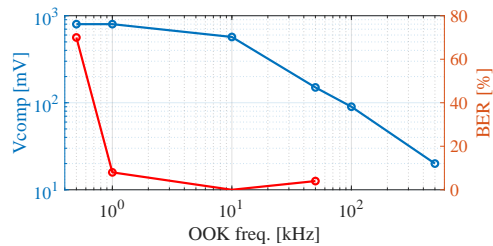}
    \caption{Bit error rate and peak-to-peak voltage before the comparator (\texttt{Vcomp}) as a function of  OOK modulation frequency.
    %As a hardware limit, for frequency larger than 40kHz no bursts can be detected (i.e. BER is not a number).
    }
    \label{fig:resber}
\end{figure}

Fig.~\ref{fig:resdelay} reports the settling time and the lost pulses as a function of the OOK modulation frequency. For the higher frequencies, the limited capabilities of the comparator are crucial. Furthermore, the ultra-low-power amplifier MAX9914, which provides the non-inverting peak-to-peak voltage (\texttt{Vcomp+}), reduces the gain as the frequency increases. The amplifier gain-bandwidth product and the comparator propagation delay lead to a longer settling time and a larger number of lost pulses at the beginning of the burst. At lower frequencies, this problem is not visible as the relaxation time of the system is smaller than the pulse period. As a final remark to further improve the analog section performance, parameters, such as the cut-off frequency of the low-pass filter, can be tailored for the specific range of OOK modulation frequencies chosen for TRAP operation to maximize the signal reception capability. As an example, the average threshold filter corner frequency may be moved accordingly for higher modulation frequencies. Thus, reducing the settling time and the number of lost pulses.

However, the system can be affected by noise appearing in the form of spurious commutations that interfere with the number of received pulses. Fig.~\ref{fig:resber} reports the bit error rate (BER) over the OOK modulation frequency and the peak-to-peak voltage before the comparator (\texttt{Vcomp}) as proof of the behaviour of the amplifier stage. While the envelop detector peak-to-peak voltage is almost constant over the frequency range, the voltage before the comparator decreases at high frequency. This situation limits the circuit to OOK modulation frequencies below 40kHz (as an upper boundary). The lower frequency boundary is mainly determined by the noise contribution, which, in our setup, has components in the range of some kHz. The noise contribution is detectable at the digital output as spurious commutations at low OOK modulation frequency disturbing the transmitted burst and the received number of pulses.

\begin{figure}
    \centering
    \includegraphics[width=0.75\columnwidth]{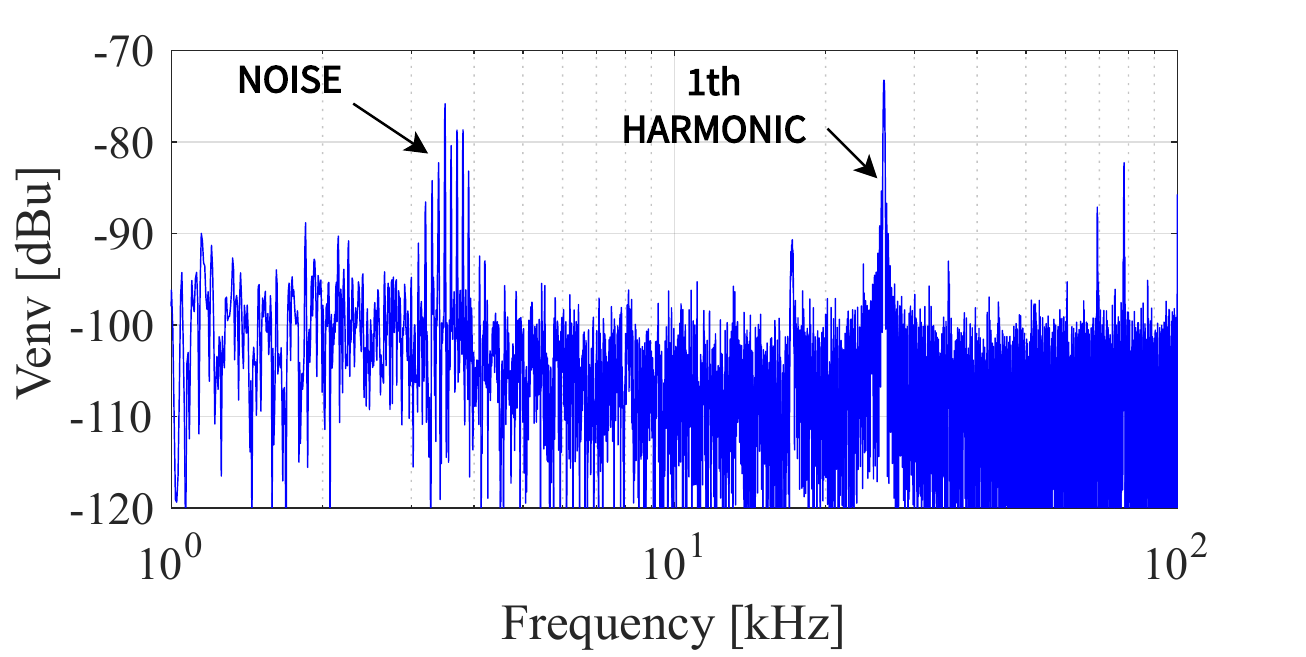}
    \caption{Frequency spectrum of the \texttt{Venv}, showing the noise at low frequencies (about 3.7kHz @ -75dBu). 
    %The first OOK modulation harmonic peak is at 26.1kHz @ -73dBu and the third harmonic is at about 78.3kHz @ -82dBu.
    }
    \label{fig:noise}
\end{figure}

Fig.~\ref{fig:noise} reports the measured envelop detector voltage \texttt{Venv} frequency spectrum where it is visible the OOK contribution with a peak at 26.1kHz for the first harmonic and 78.3kHz as third harmonic, but also, it is evident the noise contribution at lower frequencies (about 3.7kHz).

\begin{table}
\centering
\caption{Received (RX) and transmitted (TX) pulses as a function of different OOK modulation frequencies.}
\label{tab:receivedlost}
 \begin{tabular}{m{1.4cm} m{2.4cm} m{1.5cm} m{1.5cm}} 
 \hline
  {\bf TX } & {\bf OOK Freq. [kHz]} & {\bf Min RX} & {\bf Max RX} \\
 \hline
 {\bf 32} & {1.2} & {32} & {33} \\
 \rowcolor{black!3}
 {} & {12} & {32} & {32} \\
 {} & {39} & {5} & {17} \\
 \hline
 {\bf 256} & {1.2} & {256} & {257} \\
 \rowcolor{black!3}
 {} & {12} & {256} & {256} \\
 {} & {39} & {189} & {221} \\
 \hline
\end{tabular}
\end{table}

To better understand the granularity of the energy levels that the circuit can properly transmit and receive, we performed a final measurement by changing the number of transmitted pulses and the OOK modulation frequency. Table~\ref{tab:receivedlost} shows the measurement result.  At low modulation frequency (under 12kHz), the number of received pulses is almost equal to the transmitted one. Due to noise, only some more pulses are counted. At higher frequencies, due to the settling time and the limited bandwidth, the number of received pulses is smaller than the transmitted one. In particular, with a short transmitted burst (e.g., 32 or lower), the received sequence is compromised. The result confirms the upper limit at 40kHz.  However, \sysname can still operate even if some pulses are lost as it is possible to distinguish the required energy levels.

In conclusion, the circuit is fine-tuned to work in a specific range of frequencies both from the RF side (RF carrier) and OOK modulation side. A higher frequency means a higher power consumption for the oscillator but shorter on-air time for the burst transmission.% (and the possibility to scale up the system). 
The trade-off must balance these aspects and in combination with the hardware capability and noise immunity. For the \sysname implementation, we only need few energy levels (burst duration), which can be easily achieved in the evaluated range.

\begin{figure}
    \centering
   \includegraphics[width=\columnwidth]{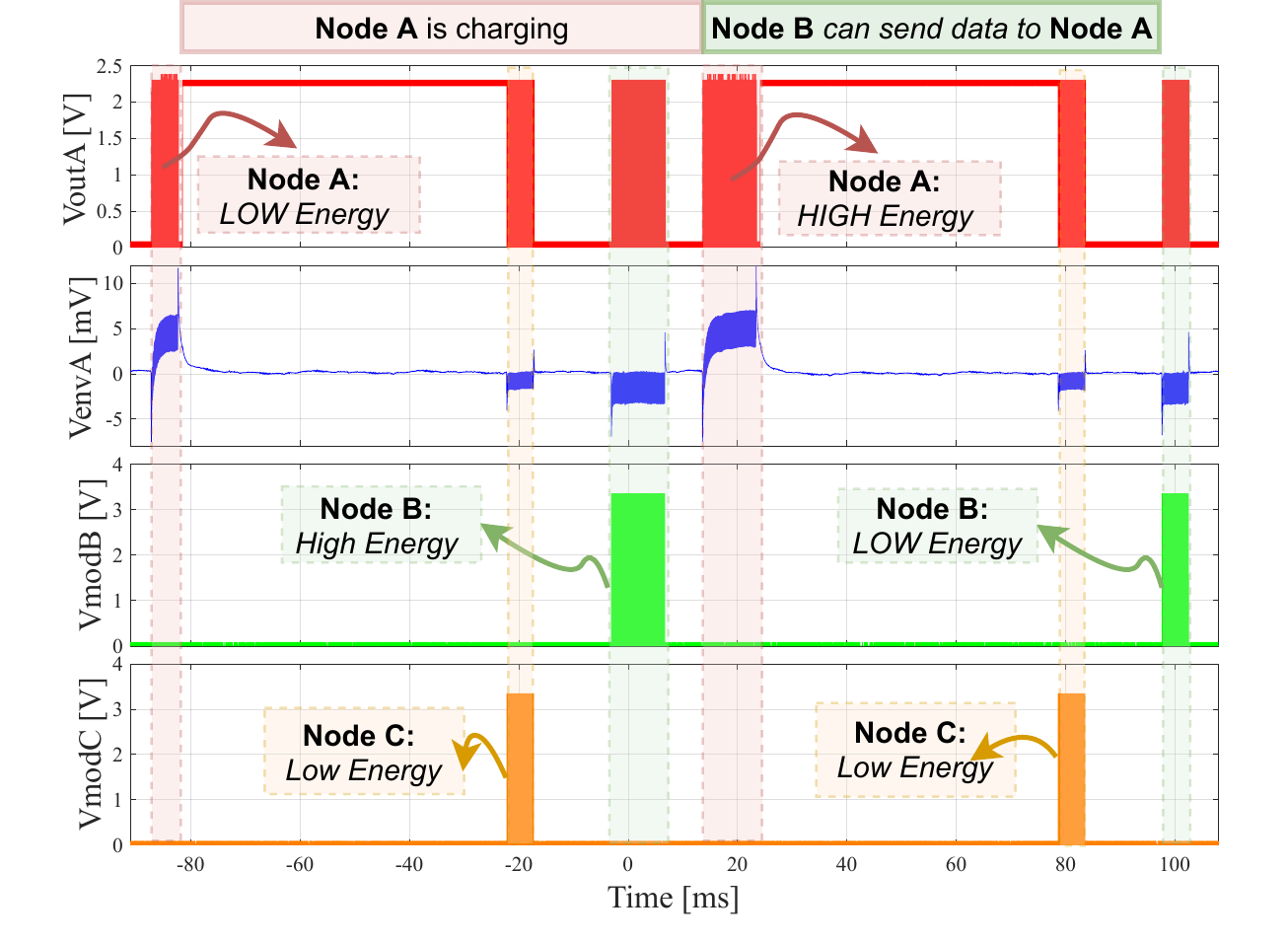}
    \caption{The most significant waveforms on the \emph{energy status channel}. Blue trace reports the envelope detector voltage of the transmitter node A (\texttt{VenvA}), and red trace depicts its digital output (\texttt{VoutA}) showing the received and demodulated bursts from the neighboring nodes. Green and orange traces refer to the modulation signals respectively at node B and node C side (\texttt{VmodB} and \texttt{VmodC}). The OOK frequency is set at 31kHz, the short burst is 128 pulses and the long burst is 256 pulses.}
    \label{fig:measuretest}
\end{figure}

\subsection{Evaluation of Energy Status Communication}

We now evaluate the energy status exchange among multiple nodes. Figure~\ref{fig:measuretest} reports the waveforms running on the test setup with three nodes. The transmitted energy status bursts from nodes B and C (\texttt{VmodB} green and \texttt{VmodC} orange traces) are used as a reference to show the behaviour of the energy status mechanism. It is worth mentioning that only two burst durations (128 and 256 pulses) are used. Node A digital output voltage (\texttt{Vout} red trace) shows the real signal that the MCU will consider understanding the energy level of the neighboring nodes B and C. The node A envelope detector voltage (\texttt{Venv} blue trace) is a good image of the signals on the energy status RF channel. They can be seen in the received and demodulated bursts from the neighboring transmitting nodes B and C concerning the burst generated by Node A (the self burst) which is much larger. Indeed, the modulator circuit on node A is directly connected to the envelope detector input. Thus, the self burst appears with a much larger variation. The scenario in the figure is summarized below:

\begin{figure}
    \centering
    \includegraphics[width=0.9\columnwidth]{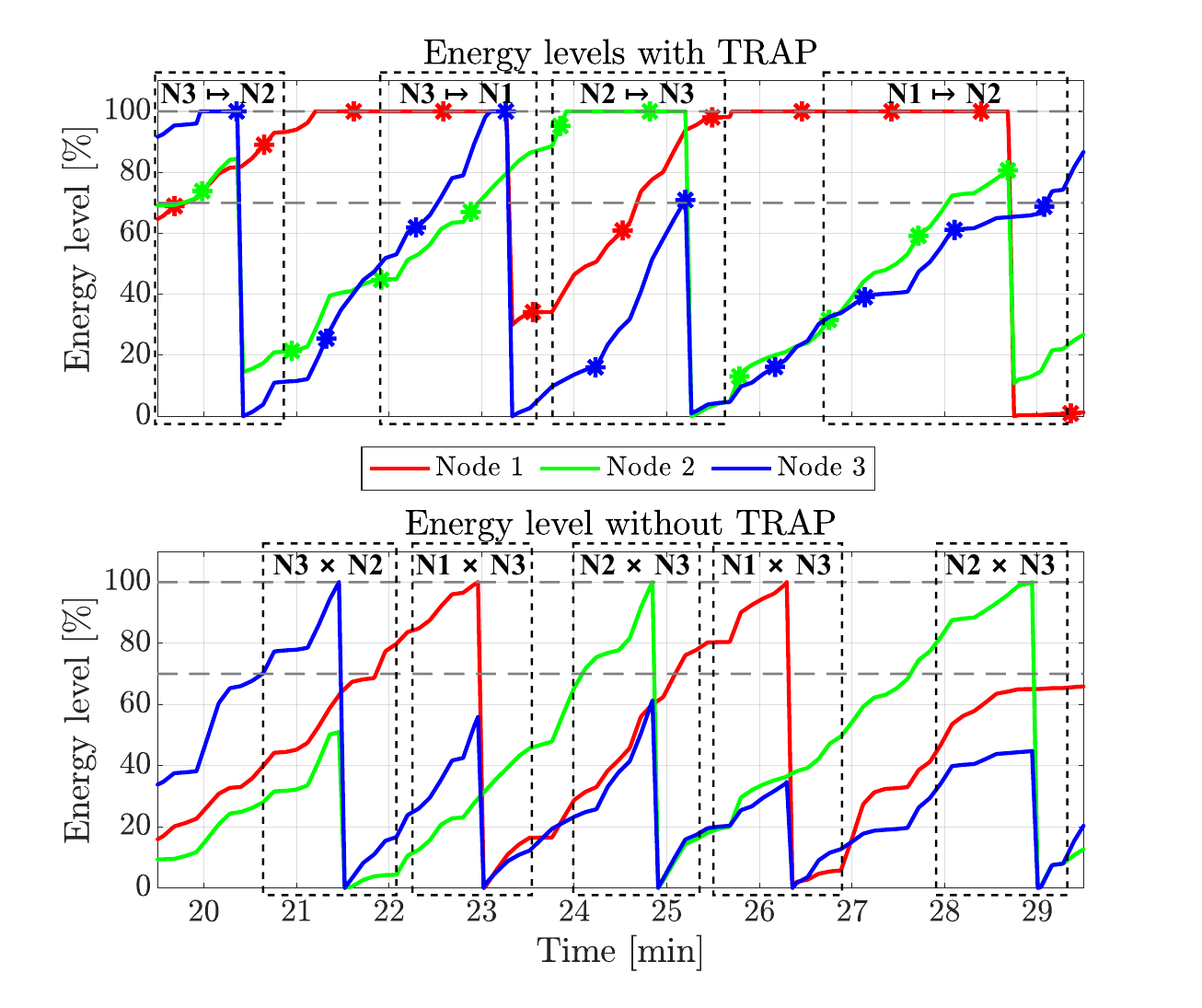}
    \label{}
    \caption{A comparison on the communication scenario with and without the TRAP implementation. Red, green, and blue traces represent the energy level of three batteryless nodes. The nodes' energy increment profile is the same for both experiments. The markers on the traces with TRAP represent the actual energy status update instant.}
    \label{fig:trap-comparison}
\end{figure}

\begin{enumerate}
    \item Node A is charging and transmits a short burst (\texttt{Venv} blue trace), communicating a lower energy availability.
    \item Node B generates a long burst (\texttt{VmodB} green trace and the respective received signal \texttt{Venv} blue trace) to show its high energy availability (i.e, able to receive data).
	\item Node C transmits short bursts indicating lower energy (\texttt{VmodC} orange trace and the respective received signal \texttt{Venv} blue trace).%, therefore it has no availability to transmit or receive sensible data.
    \item When node A has a higher energy level, it generates a long burst. node B, after the reception of the long burst from node A, can start data transmission to Node A using an active radio (or another RF channel).
    \item After the data transmission, both nodes A and B enter in lower energy status, thus transmitting short bursts.    
\end{enumerate}

\subsection{Evaluation of TRAP}

\sysname requires a software implementation (as a part of MCU firmware) to decode the energy status information. Thus, we implemented \sysname targeting STM32L476RG MCU and also a dummy application that simulates a batteryless and intermittent computing scenario, involving the characteristic charge, sense/compute/send and die cycles. Our implementation is just 76 additional lines of C code. The overhead figures are reported in Table~\ref{tab:overhead}. The additional code requires a memory overhead of just 80 bytes (i.e., the sum of data, bss, and text sections overhead). It is important to remark that from our previous version~\cite{torrisi2020zero} this overhead is minimal as the energy status transmission process is now carried out automatically by the auto-modulator circuit. Thus, meaning a higher system efficiency as the MCU can be even silent while energy status information is transmitted.
Nevertheless, MCU intervention is required to sense the backscatter channel activating the backscatter receiver circuit drawing up to 36.2$\mu$W. Only at this moment, an external GPIO interrupt is fired when a pulse (i.e., low to high transition) is received. When a node wants to transmit data to a neighbour node, it enables the external interrupt and listens to the backscatter channel. At this point, the MCU counts the number of received pulses to decode the energy status information of the neighbour node. Finally, the MCU internal clock is used to decode the burst ON-OFF keying modulation frequency.

\begin{table}
\centering
\caption{Overhead for TRAP protocol implementation.}
\label{tab:overhead}
 \begin{tabular}{ m{5cm} m{1.1cm}} 
 \hline
  {\bf Overhead} & {\bf Value} \\
 \hline
 {\em Additional lines} & 76\\
 \rowcolor{black!3}
 {\em Memory overhead [bytes]} & 80 \\
 {\em Backscatter receiver [$\mu$W]} & 36.2\\ 
 \hline
\end{tabular}
\end{table}

\begin{table}
\centering
\caption{Parameters used for the test-bed experiment}
\label{tab:trap-simulation}
 \begin{tabular}{ m{5cm} m{1.5cm}} 
 \hline
  {\bf Name} & {\bf Value} \\
 \hline
 {\em Energy required for data TX} & 100 \% \\
 \rowcolor{black!3}
 {\em Energy required for data RX} & 70 \% \\
 {\em Energy status update period} & $\sim$1 min \\
 \rowcolor{black!3}
 {\em Average energy increment} & 25 \%/min \\
 {\em Std energy increment} & 22 \%/min \\
 \hline
\end{tabular}
\end{table}

\noindpar{Testbed Setup.} We built a test-bed setup with three nodes to valuate our TRAP implementation. The previous sections presented that the backscatter channel can introduce noise, but safe burst parameters can enable an error-free energy status communication. We ran the experiment using the backscatter channel for energy status communication and a wired connection to a laptop via the UART to monitor the nodes' behavior. The nodes received roughly -20dBm signal power for the carrier. A random energy source feeds energy storage emulating the supercapacitor of a batteryless node. The energy in the supercapacitor is consumed or for a sense/compute/send action, taking all the 100\% energy, or by a receive/compute action taking 70\% energy. Thus, data transmission can happen only if the node's energy storage is fully charged, while proper data reception can happen if the energy level is above 70\%. We chose these parameters to have the possibility to test the working principles and flexibility of TRAP. Table~\ref{tab:trap-simulation} summarizes the parameter values used in our experiments.

\begin{table}
\centering
\caption{ Comparison with and without TRAP experimental result over 1 hour operation. }
\label{tab:trap-result}
 \begin{tabular}{ m{4cm} m{1.5cm}} 
 \hline
   & {\bf Total}\\
 \hline
 {\em \bf Without TRAP} &  \\
 \rowcolor{black!3}
 {\em Transmission actions} & 29 \\
 {\em Successful reception rate} & 31\% \\
 \rowcolor{black!3}
 {\em Network throughput [p/min]} & 0.15 \\
 \hline
 {\em \bf With TRAP} & \\
 \rowcolor{black!3}
 {\em Transmission actions} & 21 \\
 {\em Successful reception rate} & 100\% \\
 \rowcolor{black!3}
 {\em Network throughput [p/min]} & 0.35 \\
 \hline
\end{tabular}
\end{table}

\noindpar{Results.} We tested a communication scenario with and without TRAP by using the same energy profile. Figure~\ref{fig:trap-comparison} reports a snapshot of the behaviour of the system during the experiment where the charge, sense/compute/send, and die cycles are visible. Successful reception is established only if both transmitter and receiver have sufficient energy. Otherwise, the communication fails. Thanks to TRAP, packet losses are avoided, and all five transmission attempts (as reported in the figure succeeded). Without TRAP, all the transmissions fail due to energy unavailability on the receiver side (energy level below 70\%). However, some attempts reach a successful reception even without TRAP as, in certain situations, both sides of the data communication channel gain the proper energy level. However, one attempt reaches a successful reception even without TRAP as, in certain situations, the proper energy level is reached on both sides of the data communication channel. Table~\ref{tab:trap-result} reports the number of TX-RX actions from the network during a 1-hour experiment. The benefit of TRAP is visible as the success rate is 100\%. Without TRAP and this particular setup conditions, the packet loss is about 73\%. Moreover, it is worth mentioning that TRAP delays the data transmission until the receiver node reaches the energy level required for proper data reception. Thus, the overall TX-RX actions are less than the ones without TRAP. As the most important result, even if the attempted transmission using TRAP are lower than without TRAP, the network throughput (number of successful packet reception per time unit) with TRAP is higher as the packet loss is avoided. A remarkable result provided by our test-bed scenario shows the benefit of the presented intermittent communication protocol, which can overcome packet losses and energy waste. Finally, we reported in Figure~\ref{fig:trap-comparison} by using markers the time instant of the energy status update. Thus, showing an important aspect of our mechanism: as a receiver node transmits a mid-high level energy status (above 70\%), the transmitter immediately engages data communication. Indeed, the marker on the receiver node is just before data communication begins. We would like to emphasize that TRAP can rely on different ultra-low-power communication technologies allowing OOK modulation to share energy status information since it is independent of the physical layer (i.e., the proposed backscatter architecture).

\begin{figure}
    \centering
    \includegraphics[trim={3cm 8cm 3cm 7cm},clip, width=0.8\columnwidth]{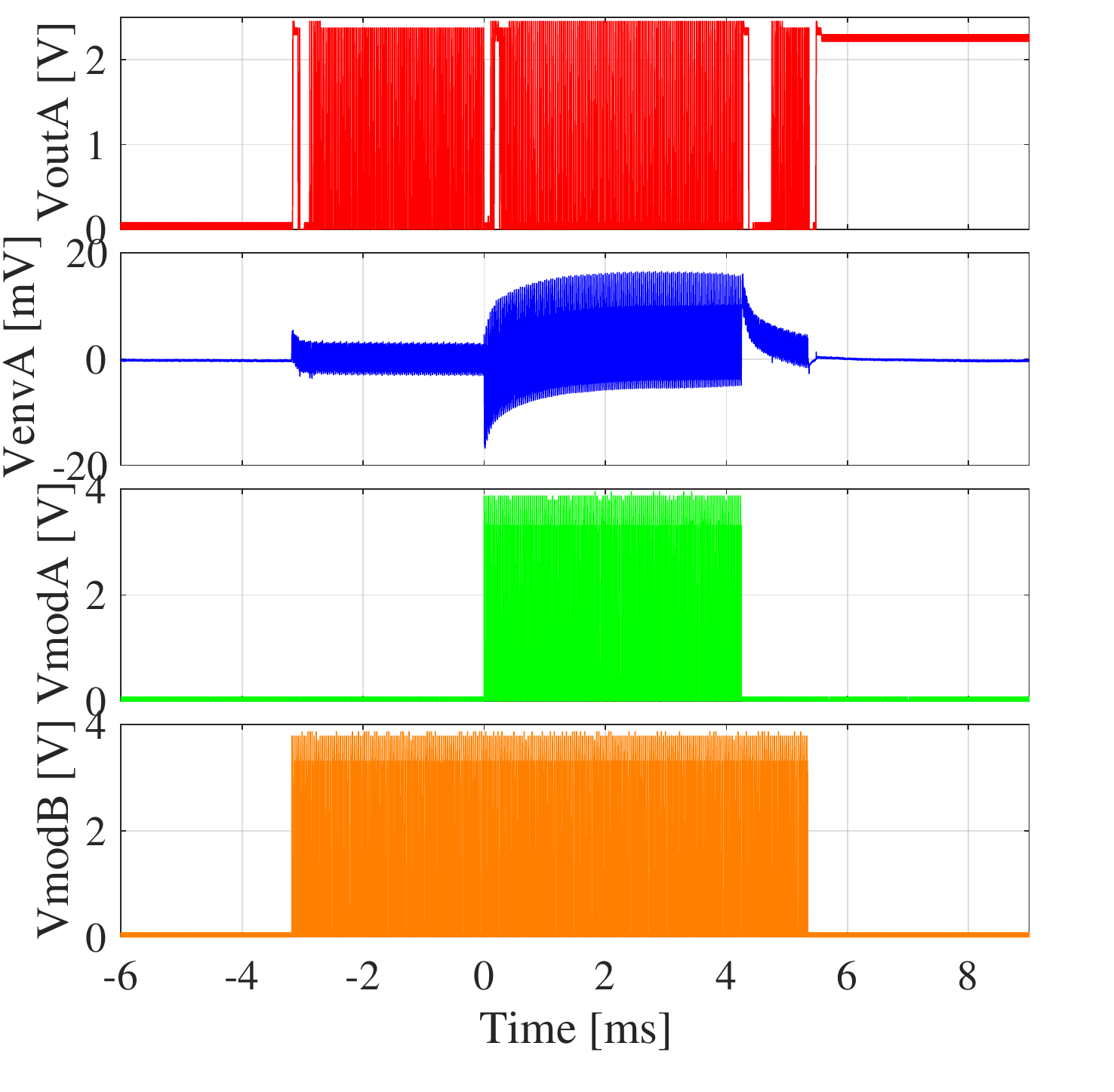}
    \caption{Evidence of the burst overlap during energy status update due to a lack of synchronization. Node A started an energy update while node B is already sending its status.}
    \label{fig:overlap}
\end{figure}

\subsection{Handling Energy Status Collisions}

Figure~\ref{fig:overlap} shows a collision event of two bursts over the backscatter channel. This situation occurs due to the nodes' local clock drift and the lack of synchronization in the energy update process. Upon energy status collision, our protocol can detect the error in the burst characteristics (i.e., number of pulses and frequency), omits the received energy status, and waits until bursts become visible again. Anyway, the receiver-initiated policy ensures energy status awareness even in this case as if the backscatter channel is blind (i.e., no burst can be properly recognized), the node waits until a clear energy update appears. Therefore, packet transmissions might delay due to collisions. Indeed, collisions manifest more likely under particular conditions related to the number of nodes, energy status update rate, and local clock drift. We leave the deeper investigation and possible hardware improvements to provide collision-free energy status transmission as future work.
\section{Conclusion}
\label{sec:conclusion}

We presented \sysname, the first intermitted communication protocol for transiently powered batteryless devices. \sysname uses an ultra-low-power backscatter circuit and a novel auto-modulator circuit that encodes the energy status over the backscatter RF channel. The energy status information is encoded by using specific frequency and duration burst, to identify both the energy level and the transmitting node. Therefore, \sysname regulates the packet transmission in transient computing devices and guarantees zero energy-wasting, since data packet transmitter nodes are aware of the energy availability of the receiver nodes. Our novelty is to provide this additional feature without any impact on the node's characteristics (such as energy consumption) since it operates autonomously. Thus, our circuit can be integrated into any existing IoT platform easily since it operates in a completely self-sustainable manner. It is also worth mentioning that our protocol is based on the transmission of energy status information over an ultra-low-power communication channel. Therefore, our protocol and the auto-modulator circuit can work with any radio that supports OOK modulation and ultra-low power requirements.
%\balance
\bibliographystyle{IEEEtran}
\bibliography{references}

\end{document}